\documentclass[twocolumn,showpacs,preprintnumbers,amsmath,amssymb,prb]{revtex4}

\usepackage{graphicx}
\usepackage{dcolumn}
\usepackage{bm}
\usepackage{tabularx}
\usepackage{amsmath}

\begin{document}

\title{Nodeless superconductivity in Ir$_{1-x}$Pt$_x$Te$_2$ with strong spin-orbital coupling}

\author{S. Y. Zhou,$^1$ X. L. Li,$^1$ B. Y. Pan,$^1$ X. Qiu,$^1$ J. Pan,$^1$ X. C. Hong,$^1$ Z. Zhang,$^1$ A. F. Fang,$^2$ N. L. Wang,$^2$ S. Y. Li$^{1,*}$}

\affiliation{$^1$State Key Laboratory of Surface Physics, Department
of Physics, and Laboratory of Advanced Materials, Fudan University,
Shanghai 200433, P. R. China\\
$^2$Beijing National Laboratory for Condensed Matter Physics, Institute of Physics, Chinese Academy of Sciences, Beijing 100190, P. R. China}

\date{\today}

\begin{abstract}
The thermal conductivity $\kappa$ of superconductor
Ir$_{1-x}$Pt$_{x}$Te$_2$ ($x$ = 0.05) single crystal with strong
spin-orbital coupling was measured down to 50 mK. The residual
linear term $\kappa_0/T$ is negligible in zero magnetic field. In
low magnetic field, $\kappa_0/T$ shows a slow field dependence.
These results demonstrate that the superconducting gap of
Ir$_{1-x}$Pt$_{x}$Te$_2$ is nodeless, and the pairing symmetry is
likely conventional $s$-wave, despite the existence of strong
spin-orbital coupling and a quantum critical point.

\end{abstract}


\maketitle

\section{Introduction}

The effect of strong spin-orbital coupling (SOC) on
superconductivity has recently attracted much attention. One example
is the topological superconductor, such as candidate
Cu$_x$Bi$_2$Se$_3$ in which Cu atoms are intercalated into
topological insulator Bi$_2$Se$_3$ with strong SOC. \cite{YSHor}
Novel superconducting state was claimed in Cu$_x$Bi$_2$Se$_3$ by the
point-contact spectra and superfluid density measurements.
\cite{SSasaki,MKriener1} Another example is the noncentrosymmetric
superconductor, such as Li$_2$Pt$_3$B in which the spatial inversion
symmetry is broken. \cite{PBadica} The strong SOC in Li$_2$Pt$_3$B
gives large spin-triplet pairing component and produces line nodes
in the superconducting gap. \cite{HQYuan,MNishiyama}

More recently, superconductivity was discovered in the layered
compound IrTe$_2$ by Pd intercalation (Pd$_x$IrTe$_2$),
\cite{JJYang} Pd substitution (Ir$_{1-x}$Pd$_x$Te$_2$),
\cite{JJYang} Pt substitution (Ir$_{1-x}$Pt$_x$Te$_2$), \cite{SPyon}
or Cu intercalation (Cu$_x$IrTe$_2$). \cite{MKamitani} Since the SOC
is proportional to $Z^4$, where $Z$ is the atomic number, the
superconductivity in doped IrTe$_2$ must associated with strong SOC
due to the large $Z$.

Furthermore, the parent compound IrTe$_2$ exhibits an intriguing
structural phase transition from a high-temperature trigonal to a
low-temperature monoclinic phase near 270 K. \cite{JJYang} Initially
it was related to a charge-density-wave (CDW) induced by Ir $5d$
$t_{2g}$ orbitals, \cite{JJYang,DOotsuki} however, later no CDW gap
was detected from the optical spectroscopy \cite{AFFang} and
angle-resolved photoemission spectroscopy (ARPES)
\cite{DOotsukiARPES} measurements. At this moment, the origin of the
transition is still under hot debate, with proposals such as crystal
field effect from Te $5p$ orbital splitting, \cite{AFFang} the
depolymerization-polymerization of anionic Te bonds, \cite{YSOhPRL}
and Ir $5d$ orbital order. \cite{HBCao} With increasing the doping
level $x$, the structure transition is gradually suppressed and
superconductivity emerges, showing a dome-like phase diagram with
the maximum $T_c$ of 3 K near $x \approx$ 0.04. \cite{JJYang,SPyon}
Such a phase diagram of doped IrTe$_2$ is reminiscent of high-$T_c$
cuprates and some heavy fermion superconductors, in which
superconductivity appears close to a magnetic quantum critical point
(QCP). This means that there likely exists a QCP under the
superconducting dome of doped IrTe$_2$, and the superconductivity
may be unconventional. \cite{PMonthoux} Therefore it is of great
interest to investigate whether there is novel superconducting state
in doped IrTe$_2$.

The ultra-low-temperature thermal conductivity measurement is a bulk
tool to study the gap structure of superconductors.
\cite{Shakeripour} The existence of a finite residual linear term
$\kappa_0/T$ in zero field is usually considered as the signature of
nodal superconducting gap. Further information of nodal gap, gap
anisotropy, or multiple gaps may be obtained from the field
dependence of $\kappa_0/T$. \cite{Shakeripour} Previously,
single-gap $s$-wave superconductivity near the QCP of CDW has
been clearly shown in Cu$_x$TiSe$_2$ by thermal conductivity
measurements. \cite{SYLi1}

In this paper, we probe the superconducting gap structure of
Ir$_{1-x}$Pt$_{x}$Te$_2$ ($x$ = 0.05) single crystal by measuring
the thermal conductivity $\kappa$ down to 50 mK. The residual linear
term $\kappa_0/T$ is negligible in zero magnetic field. The field
dependence of $\kappa_0/T$ is slow at low field, unlike that of a
nodal superconductor. Both results suggest nodeless superconducting
gap in Ir$_{1-x}$Pt$_{x}$Te$_2$.

\section{Experimental}

Single crystals of Ir$_{1-x}$Pt$_x$Te$_2$ were grown via self-flux
method. \cite{AFFang} The dc magnetic susceptibility was measured by
using a SQUID (MPMS, Quantum Design). The heat capacity measurement
was preformed in a physical property measurement system (PPMS,
Quantum Design) via the relaxation method. The
Ir$_{0.95}$Pt$_{0.05}$Te$_2$ single crystal was cut to a rectangular
shape of dimensions 2.0 $\times$ 0.55 mm$^2$ in the $ab$ plane and
20 $\mu$m thickness along the $c$ axis. Four silver wires were
attached to the sample surface with silver paint, which were used
for both in-plane resistivity and thermal conductivity measurements.
The contacts are metallic with typical resistance 15 m$\Omega$ at 2
K. In-plane thermal conductivity was measured in a dilution
refrigerator, using a standard four-wire steady-state method with
two RuO$_2$ chip thermometers, calibrated {\it in situ} against a
reference RuO$_2$ thermometer. Magnetic fields were applied along
the $c$ axis and perpendicular to the heat current. To ensure a
homogeneous field distribution in the sample, all fields for
resistivity and thermal conductivity measurements were applied at
temperature above $T_c$.

\begin{figure}
\includegraphics[clip,width=7cm]{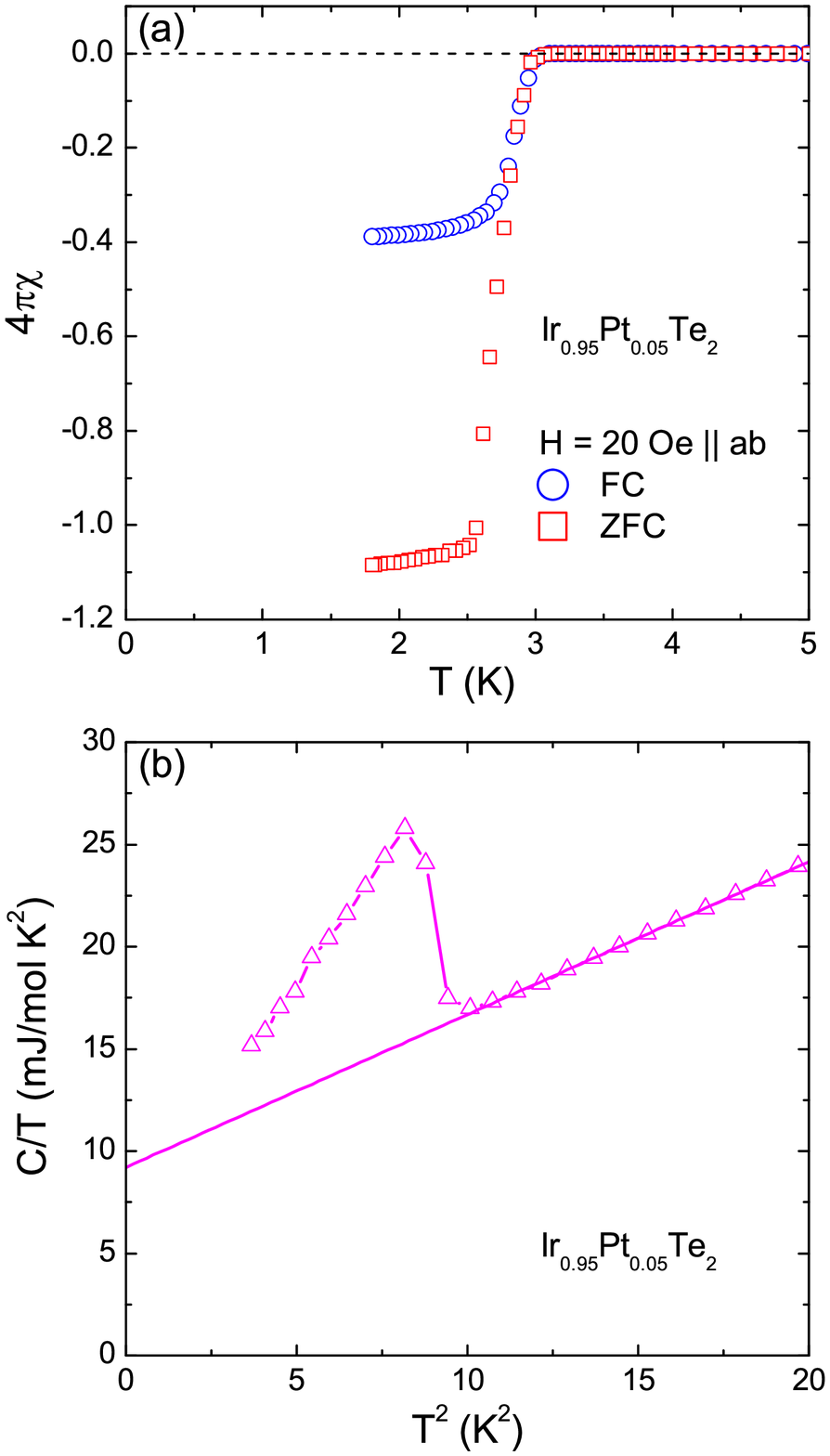}
\caption{(Color online). (a) Low-temperature magnetic susceptibility
of Ir$_{0.95}$Pt$_{0.05}$Te$_2$ single crystal. The measurements
were performed in an applied field of $H$ = 20 Oe parallel to the
$ab$ plane with zero-field-cooled (ZFC) and field-cooled (FC)
processes. (b) Specific heat $C/T$ as a function of $T^2$ for
Ir$_{0.95}$Pt$_{0.05}$Te$_2$ single crystals. The $C/T$ versus $T^2$
shows a linear dependence above $T_c$, giving the electronic
specific-heat coefficient $\gamma$ = 9.20 mJ/mol K$^2$. }
\end{figure}

\begin{figure}
\includegraphics[clip,width=9cm]{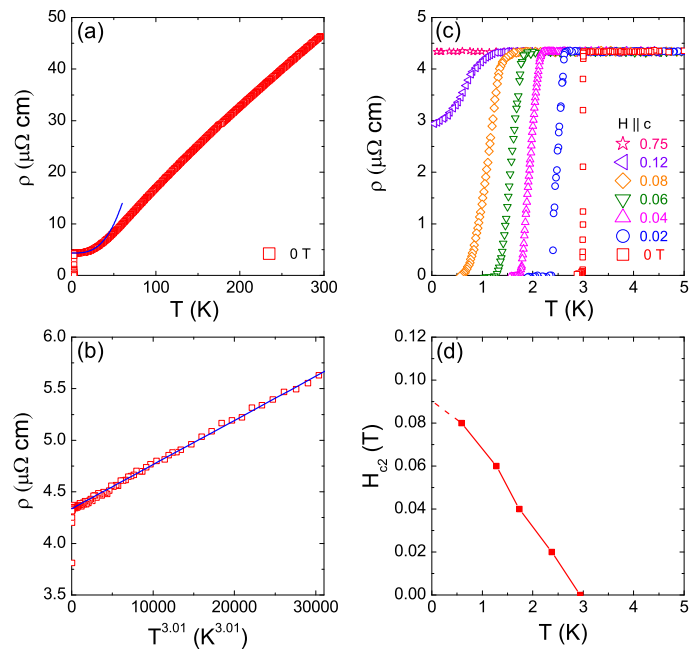}
\caption{(Color online). (a) Temperature dependence of the
resistivity $\rho$ for Ir$_{0.95}$Pt$_{0.05}$Te$_2$ single crystal.
The data between 3.5 and 31 K can be fitted to $\rho(T) = \rho_0
+AT^n$, as shown by the solid line, with $\rho_0$ = 4.34 $\pm$ 0.002
$\mu\Omega$cm and $n$ = 3.01 $\pm$ 0.03. (b) The resistivity $\rho$
as a function of $T^{3.01}$ below 31 K. The solid line is the
fitting curve in (a). (c) Low-temperature resistivity of
Ir$_{0.95}$Pt$_{0.05}$Te$_2$ single crystal in magnetic fields up to
0.75 T. (d) Temperature dependence of the upper critical field
$H_{c2}$, defined at the point $\rho$ dropping to zero on the curves
in (c). The dashed line is a guide to the eye, which points to
$H_{c2}(0) \approx$ 0.09 T.}
\end{figure}

\section{Results and Discussion}

Figure 1(a) presents the dc magnetic susceptibility of
Ir$_{0.95}$Pt$_{0.05}$Te$_2$ single crystal. It was measured in
magnetic field $H$ = 20 Oe parallel to the $ab$ plane, with
zero-field-cooled (ZFC) and field-cooled (FC) processes. Sharp
superconducting transition with $T_c \approx$ 3.0 K and about 100\%
shielding volume fraction were observed for the ZFC process,
suggesting the homogeneous bulk superconductivity in our sample.

Heat capacity was measured on three pieces of
Ir$_{0.95}$Pt$_{0.05}$Te$_2$ single crystals, with a total mass of
8.8 mg. It is plotted in Fig. 1(b), as $C/T$ versus $T^2$. Above
$T_c$, the data can be well fitted by $C/T = \gamma +\beta T^2$,
giving the electronic specific-heat coefficient $\gamma$ = 9.20
mJ/mol K$^2$. The significant jump was observed at $T_c \approx$ 3.0
K, which also indicates the high quality of our single crystals.

Figure 2(a) shows the resistivity of Ir$_{0.95}$Pt$_{0.05}$Te$_2$
single crystal in zero field. No resistivity anomaly is observed
above $T_c$, suggesting that no structural transition occurs in this
sample near optimal doping. The data between 3.5 and 31 K can be
fitted to $\rho(T) = \rho_0 +AT^n$, with $\rho_0$ = 4.34 $\pm$ 0.002
$\mu\Omega$cm and $n$ = 3.01 $\pm$ 0.03. To see more clearly, Fig.
2(b) plots $\rho$ versus $T^{3.01}$ and the solid line represents
the fitting curve data in Fig. 2(a). Such a temperature dependence
of $\rho(T) \sim T^n$ with $n \approx 2.8$ has been observed in
Ir$_{1-x}$Pt$_x$Te$_2$ polycrystal and attributed to phonon-assisted
interband scattering, \cite{SPyon} as in TiSe$_2$.
\cite{AFKusmartseva}

Previously, the upper critical field $H_{c2}(0) \approx$ 0.17 T has
been determined for Ir$_{0.96}$Pt$_{0.04}$Te$_2$ polycrystal by
resistivity measurements. \cite{SPyon} In order to obtain the
$H_{c2}(0)$ of our Ir$_{0.95}$Pt$_{0.05}$Te$_2$ single crystal, we
also measure its resistivity with magnetic field parallel to the $c$
axis up to $H$ = 0.75 T and down to 50 mK, shown in Fig. 2(c). In
zero field, the resistivity drops to zero at 2.94 K with a narrow
transition width of 0.06 K. The superconducting transition is
gradually suppressed in magnetic fields. In Fig. 2(d), we plot the
temperature dependence of $H_{c2}$, where $T_c$ is defined at the
resistivity $\rho$ dropping to zero on the curves in Fig. 2(c). The
dashed line is a guide to the eye, which points to $H_{c2}(0)
\approx$ 0.09 T. This value is only about half of that in
Ir$_{0.96}$Pt$_{0.04}$Te$_2$ polycrystal. \cite{SPyon} It is not
surprising since in polycrystal, due to the random grain
orientation, the $H_{c2}$ represents the maximum value for all field
configurations, presumably $H \parallel ab$.

\begin{figure}
\includegraphics[clip,width=7.5cm]{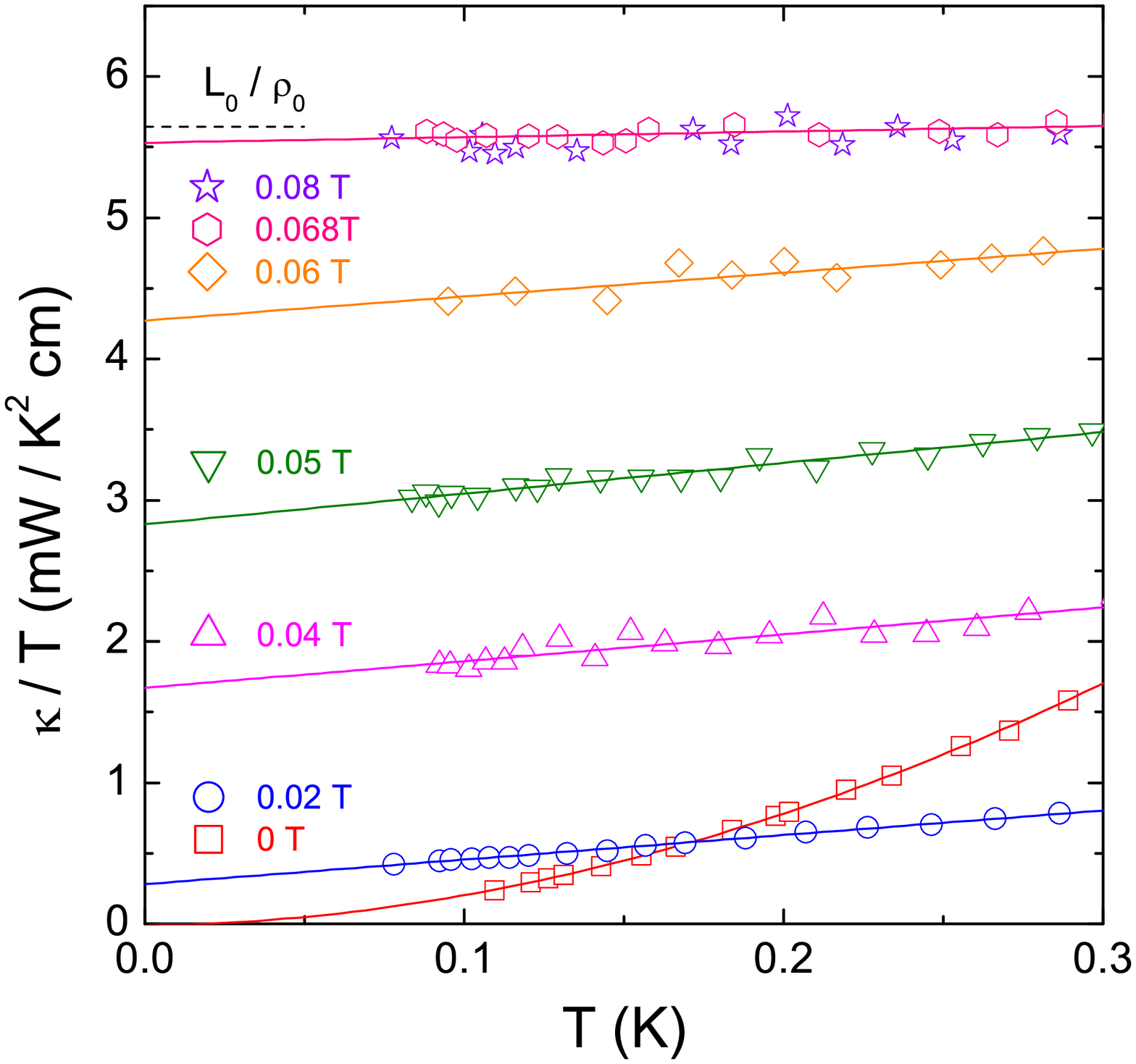}
\caption{(Color online). Low-temperature thermal conductivity of
Ir$_{0.95}$Pt$_{0.05}$Te$_2$ single crystal in zero and magnetic
fields. The solid lines are fits to $\kappa/T = a + bT^{\alpha-1}$.
The dash line is the normal-state Wiedemann-Franz law expectation
$L_0$/$\rho_0$, with the Lorenz number $L_0$ = 2.45 $\times$
10$^{-8}$ W$\Omega$K$^{-2}$ and $\rho_0$ = 4.34 $\mu\Omega$cm.}
\end{figure}

\begin{figure}
\includegraphics[clip,width=7.73cm]{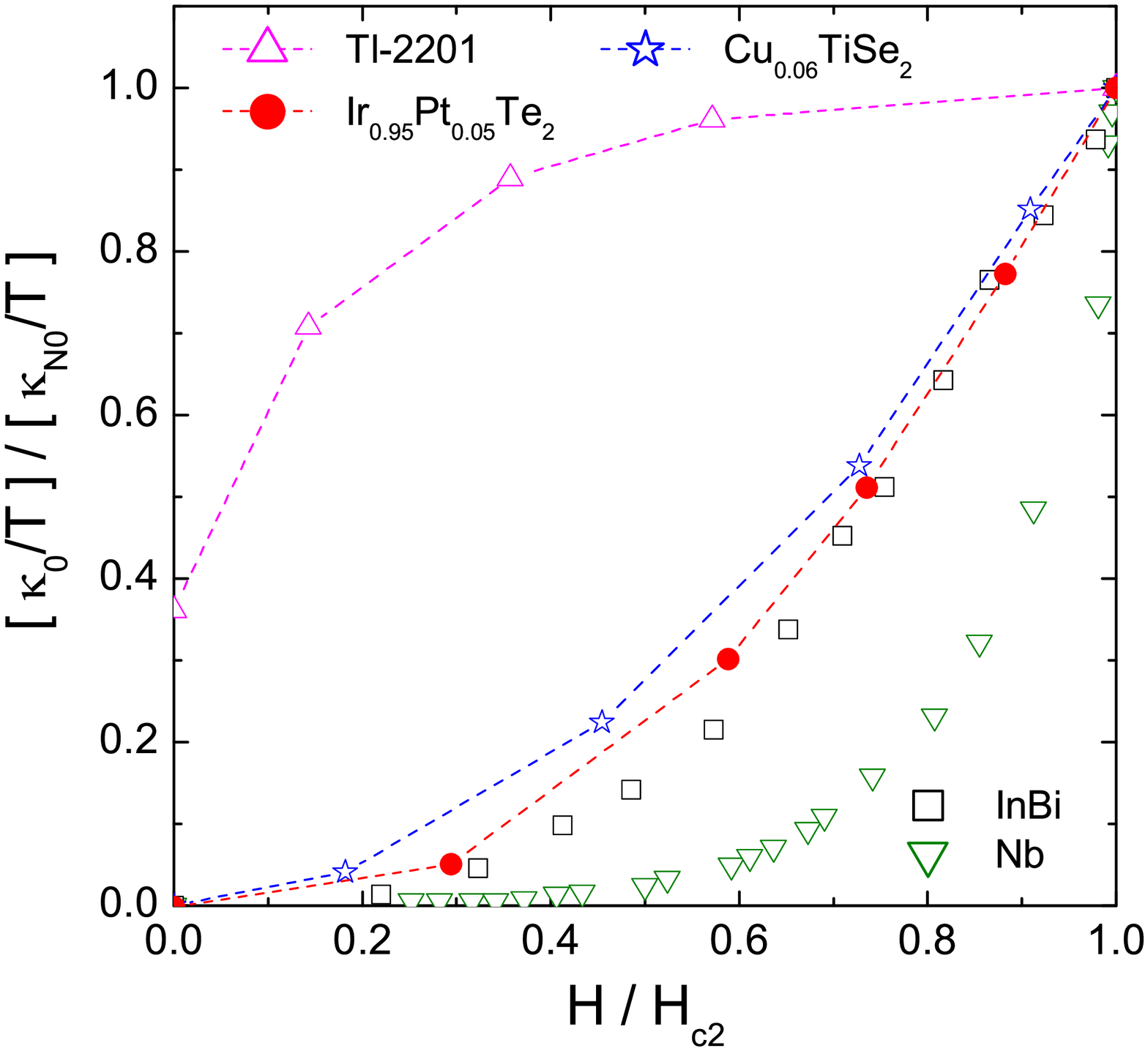}
\caption{(Color online). Normalized residual linear term
$\kappa_0/T$ of Ir$_{0.95}$Pt$_{0.05}$Te$_2$ as a function of
$H/H_{c2}$. For comparison, similar data are shown for the clean
$s$-wave superconductor Nb, \cite{Lowell} the dirty $s$-wave
superconducting alloy InBi, \cite{Willis} an overdoped $d$-wave
cuprate superconductor Tl-2201, \cite{Proust} and the single-gap
$s$-wave superconductor Cu$_{0.06}$TiSe$_2$. \cite{SYLi1}}
\end{figure}

The thermal conductivities of Ir$_{0.95}$Pt$_{0.05}$Te$_2$ single
crystal in zero and magnetic fields up to $H$ = 0.08 T are plotted
in Fig. 3, as $\kappa / T$ versus $T$. We fit all the curves to
$\kappa/T$ = $a + bT^{\alpha-1}$, in which the two terms $aT$ and
$bT^{\alpha}$ represent contributions from electrons and phonons,
respectively. \cite{Sutherland1,SYLi2} The power $\alpha$ of the
second term contributed by phonons is typically between 2 and 3 for
single crystals, due to the specular reflections of phonons at the
boundary. \cite{Sutherland1,SYLi2} In zero field, the fitting gives
a residual linear term $\kappa_0/T \equiv a$ = -7 $\pm$ 18 $\mu$W
K$^{-2}$ cm$^{-1}$, with $\alpha$ = 2.91 $\pm$ 0.04. Since our
experimental error bar is 5 $\mu$W K$^{-2}$cm$^{-1}$, the
$\kappa_0/T$ in zero filed is essentially zero, comparing to the
normal-state Wiedemann-Franz law expectation $L_0$/$\rho_0$ = 5.65
mW K$^{-2}$ cm$^{-1}$, with the Lorenz number $L_0$ = 2.45 $\times$
10$^{-8}$ W$\Omega$K$^{-2}$ and $\rho_0$ = 4.34 $\mu\Omega$cm.

For $s$-wave nodeless superconductors, there are no fermionic
quasiparticles to conduct heat when $T \rightarrow 0$ since all
electrons become Cooper pairs. Therefore, there is no residual
linear term of $\kappa_0/T$, as seen in Nb \cite{Lowell} and InBi.
\cite{Willis} However, a finite $\kappa_0/T$ in zero field is
usually observed in a superconductor with nodal gap, coming from the
nodal quasiparticles. \cite{Shakeripour} For example, $\kappa_0/T$ =
1.41 mW K$^{-2}$ cm$^{-1}$ for the overdoped cuprate
Tl$_2$Ba$_2$CuO$_{6+\delta}$ (Tl-2201), a $d$-wave superconductor
with $T_c$ = 15 K, \cite{Proust} and $\kappa_0/T$ = 17 mW K$^{-2}$
cm$^{-1}$ for the ruthenate Sr$_2$RuO$_4$, a $p$-wave superconductor
with $T_c$ = 1.5 K. \cite{Suzuki} The negligible $\kappa_0/T$ of our
Ir$_{0.95}$Pt$_{0.05}$Te$_2$ single crystal in zero field suggests a
nodeless superconducting gap.

The field dependence of $\kappa_0/T$ will give more information of
the superconducting gap structure. For a clean type-II $s$-wave
superconductor with isotropic gap, $\kappa_0/T$ should grow
exponentially with field (above $H_{c1}$), as in Nb. \cite{Lowell}
In the case of nodal superconductor, $\kappa_0/T$ increases rapidly
($\sim H^{1/2}$) in low field due to the Volovik effect,
\cite{Volovik} as in Tl-2201. \cite{Proust}

In Fig. 3, all the curves in magnetic field are roughly linear.
Therefore we fit these curves to $\kappa/T$ = $a + bT^{\alpha-1}$
with $\alpha$ fixed to 2. Previously $\alpha \approx$ 2.2 was found
in Cu$_{0.06}$TiSe$_2$, \cite{SYLi1} and recently $\alpha \approx$ 2
has been observed in some iron-based superconductors such as
BaFe$_{1.9}$Ni$_{0.1}$As$_2$, \cite{LDing} KFe$_2$As$_2$,
\cite{JKDong} and Ba(Fe$_{1-x}$Ru$_x$)$_2$As$_2$ single crystals.
\cite{XQiu} Note that in the field of $H$ = 0.068 T and 0.08 T,
$\kappa_0/T$ = 5.53 $\pm$ 0.03 and 5.49 $\pm$ 0.05 mW K$^{-2}$
cm$^{-1}$ were obtained, respectively. Both values are close to the
normal-state Wiedemann-Franz law expectation $L_0$/$\rho_0$ = 5.65
mW K$^{-2}$ cm$^{-1}$. We take $H$ = 0.068 T as its bulk
$H_{c2}(0)$. A slightly different $H_{c2}(0)$ does not affect our
discussion on the field dependence of $\kappa_0/T$ below.

In Fig. 4, the normalized $\kappa_0/T$ of
Ir$_{0.95}$Pt$_{0.05}$Te$_2$ single crystal is plotted as a function
of $H/H_{c2}$. For comparison, we also plot the data of the clean
$s$-wave superconductor Nb, \cite{Lowell} the dirty $s$-wave
superconducting alloy InBi, \cite{Willis} the $d$-wave cuprate
superconductor Tl-2201, \cite{Proust} and the $s$-wave
superconductor Cu$_{0.06}$TiSe$_2$. \cite{SYLi1} From Fig. 4, the
$\kappa_0/T$ of Ir$_{0.95}$Pt$_{0.05}$Te$_2$ shows a field
dependence similar to that of $s$-wave superconductors InBi and
Cu$_{0.06}$TiSe$_2$. This further supports that
Ir$_{0.95}$Pt$_{0.05}$Te$_2$ is a nodeless superconductor.

Nodeless gap does not essentially mean conventional
superconductivity. For example, the nodeless gap observed in
optimally doped iron-based superconductor may be unconventional
$s_{\pm}$-wave resulting from antiferromagnetic spin fluctuations.
\cite{Hirschfeld} However, here in Ir$_{1-x}$Pt$_x$Te$_2$, the
nodeless gap is unlikely $s_{\pm}$-wave. One obvious reason is that
Ir$_{1-x}$Pt$_x$Te$_2$ does not have that kind of multiple Fermi
surfaces as in iron-based superconductors. \cite{JJYang,Hirschfeld}
Therefore, the pairing symmetry in Ir$_{1-x}$Pt$_x$Te$_2$ is likely
conventional $s$-wave. In this sense, the appearance of
superconductivity may have little relationship with quantum
fluctuations near QCP. This is not surprising since the single-gap
$s$-wave superconductivity in Cu$_x$TiSe$_2$ is also not related to
quantum fluctuations of CDW.

From the aspect of strong SOC, an unconventional odd-parity pairing
state was claimed for the topological superconductor candidate
Cu$_x$Bi$_2$Se$_3$. \cite{SSasaki,MKriener1} Both nodeless gap or
gap with point nodes are allowed for the odd-parity superconducting
state. \cite{SSasaki} In this context, more experiments such as
point-contact spectra are needed to completely exclude novel
superconductivity in Ir$_{1-x}$Pt$_x$Te$_2$.

\section{Summary}

In summary, we investigate the superconducting gap structure of
Ir$_{1-x}$Pt$_{x}$Te$_2$ ($x$ = 0.05) single crystal by thermal
conductivity measurements. The $\kappa_0/T$ in zero field is
negligible, and the field dependence of $\kappa_0/T$ is slow at low
field. Both of them suggest nodeless superconductivity in
Ir$_{1-x}$Pt$_{x}$Te$_2$. The pairing symmetry is likely
conventional $s$-wave, although the odd-parity superconducting state
can not be completely excluded from our measurements.

\begin{center}
{\bf ACKNOWLEDGEMENTS}
\end{center}
We thank J. J. Yang, Y. S. Oh, and S-W. Cheong for providing
Ir$_{0.96}$Pt$_{0.04}$Te$_2$ polycrystal and nominal
Ir$_{0.8}$Pt$_{0.2}$Te$_2$ single crystal to initialize this study.
This work is supported by the Natural Science Foundation of China,
the Ministry of Science and Technology of China (National Basic
Research Program No: 2009CB929203 and 2012CB821402), and the Program
for Professor of Special Appointment (Eastern Scholar) at Shanghai
Institutions of Higher Learning. \\

$^*$ E-mail: shiyan$\_$li@fudan.edu.cn

\end{document}